\documentstyle[aaspp4]{article}	
\makeatletter
\let\@internalcite\cite
\def\cite{\def\@citeseppen{-1000}%
    \def\@cite##1##2{(##1\if@tempswa , ##2\fi)}%
    \def\citeauthoryear##1##2##3{##1 ##3}\@internalcite}

\def\citeNP{\def\@citeseppen{-1000}%
    \def\@cite##1##2{##1\if@tempswa , ##2\fi}%
    \def\citeauthoryear##1##2##3{##1 ##3}\@internalcite}
\def\citeN{\def\@citeseppen{-1000}%
    \def\@cite##1##2{##1\if@tempswa , ##2)\else{)}\fi}%
    \def\citeauthoryear##1##2##3{##1 (##3}\@citedata}
\def\citeA{\def\@citeseppen{-1000}%
    \def\@cite##1##2{(##1\if@tempswa , ##2\fi)}%
    \def\citeauthoryear##1##2##3{##1}\@internalcite}
\def\citeANP{\def\@citeseppen{-1000}%
    \def\@cite##1##2{##1\if@tempswa , ##2\fi}%
    \def\citeauthoryear##1##2##3{##1}\@internalcite}
\def\shortcite{\def\@citeseppen{-1000}%
    \def\@cite##1##2{(##1\if@tempswa , ##2\fi)}%
    \def\citeauthoryear##1##2##3{##2 ##3}\@internalcite}
\def\shortciteNP{\def\@citeseppen{-1000}%
    \def\@cite##1##2{##1\if@tempswa , ##2\fi}%
    \def\citeauthoryear##1##2##3{##2 ##3}\@internalcite}
\def\shortciteN{\def\@citeseppen{-1000}%
    \def\@cite##1##2{##1\if@tempswa , ##2)\else{)}\fi}%
    \def\citeauthoryear##1##2##3{##2 (##3}\@citedata}
\def\shortciteA{\def\@citeseppen{-1000}%
    \def\@cite##1##2{(##1\if@tempswa , ##2\fi)}%
    \def\citeauthoryear##1##2##3{##2}\@internalcite}
\def\shortciteANP{\def\@citeseppen{-1000}%
    \def\@cite##1##2{##1\if@tempswa , ##2\fi}%
    \def\citeauthoryear##1##2##3{##2}\@internalcite}
\def\citeyear{\def\@citeseppen{-1000}%
    \def\@cite##1##2{(##1\if@tempswa , ##2\fi)}%
    \def\citeauthoryear##1##2##3{##3}\@citedata}
\def\citeyearNP{\def\@citeseppen{-1000}%
    \def\@cite##1##2{##1\if@tempswa , ##2\fi}%
    \def\citeauthoryear##1##2##3{##3}\@citedata}

\def\@citedata{%
	\@ifnextchar [{\@tempswatrue\@citedatax}%
				  {\@tempswafalse\@citedatax[]}%
}

\def\@citedatax[#1]#2{%
\if@filesw\immediate\write\@auxout{\string\citation{#2}}\fi%
  \def\@citea{}\@cite{\@for\@citeb:=#2\do%
    {\@citea\def\@citea{, }\@ifundefined
       {b@\@citeb}{{\bf ?}%
       \@warning{Citation `\@citeb' on page \thepage \space undefined}}%
{\csname b@\@citeb\endcsname}}}{#1}}%

\def\@citex[#1]#2{%
\if@filesw\immediate\write\@auxout{\string\citation{#2}}\fi%
  \def\@citea{}\@cite{\@for\@citeb:=#2\do%
    {\@citea\def\@citea{; }\@ifundefined
       {b@\@citeb}{{\bf ?}%
       \@warning{Citation `\@citeb' on page \thepage \space undefined}}%
{\csname b@\@citeb\endcsname}}}{#1}}%

\def\@biblabel#1{}

\newlength{\bibhang}
\makeatother

\newcommand{\fig}[1]{Fig.~\ref{#1}}

\def\htwo{H$_2$}

\def\orf{ORFEUS}

\begin{document}
 
\title{Far Ultraviolet Fluorescence of Molecular Hydrogen\\
in IC 63\footnotemark}
\footnotetext{Based on the development and utilization of
\orf\ (Orbiting and Retrievable Far and Extreme Ultraviolet
Spectrometers), a collaboration of the Astronomical Institute of the
University of T\"{u}bingen, the Space Astrophysics Group of the
University of California at Berkeley, and the Landessternwarte
Heidelberg.}


\author{Mark Hurwitz}
\affil{Space Sciences Laboratory\\
University of California, Berkeley, California 94720-5030}
\authoremail{markh@ssl.berkeley.edu}


\begin{abstract}

We present observations of \htwo\ fluorescence at
wavelengths between 1000 and 1200 \AA\ from the bright reflection nebula IC 63.
Observations were performed with the Berkeley
spectrograph aboard the \orf-II mission \cite{Hetal98}.
To the best of our knowledge, this is the first
detection of astrophysical \htwo\ fluorescent
emission at these wavelengths (excluding planetary atmospheres).
The shape of the spectrum is well described by the model of \citeN{S89}.
The absolute intensity, however, is fainter than an 
extrapolation from observations 
at longer ultraviolet wavelengths \cite{WSBB89}
by a factor of ten.  Of the mechanisms that might help
reconcile these observations, optical depth effects in
the fluorescing \htwo\ itself are the most promising
(or at least the most difficult to rule out).

\end{abstract}
 

\section{INTRODUCTION}

In many environments, the equilibrium abundance of the hydrogen 
molecule (\htwo) is determined 
by a balance between formation on dust grains and
photodissociation pumped by photons at wavelengths
below about 1108 \AA\ \cite{SB82}.
Photons between this threshold energy and
the Lyman limit populate a large number 
of rovibrational states in the $B^1 \Sigma_u^+$ and $C^1 \Pi_u$ electronic levels.
En route back to the $X^1 \Sigma_g^+$ level,
the excited \htwo\ may radiate through 
any of several bound-bound and/or bound-free channels
leading to a complex fluorescence emission spectrum.

Located about 20$\arcmin$ or 1.3 pc from $\gamma$ Cas, 
dense gas in the reflection nebula IC 63 is illuminated by a
bright UV radiation field.  Line widths of a variety
of molecular species are quite narrow \cite{JVDB94},
suggesting that shock excitation is comparatively unimportant
in this environment.  The nebula provides an excellent test case for
models of \htwo\ fluorescence and other
photochemical processes \cite{JVDBSS95}.

The fluorescence model described in \citeN{S88} and \citeN{S89} has been
successfully applied in the interpretation of 
infrared \htwo\ fluorescence from the reflection nebulae
IC 63 and IC 59 \cite{LLBJF97},
and to the ultraviolet fluorescence
in the band near 1600 \AA\ 
from IC 63 \cite{S89} and the
Taurus molecular cloud \cite{H94}.
In this work we test the model in a previously
unexplored band deep in the ultraviolet
where the significant majority of the radiated fluorescent
power is expected to emerge.

\section{OBSERVATIONS AND DATA REDUCTION}
\label{reduction}

The Berkeley EUV/FUV spectrometer, located at the prime focus of the
1-m \orf\ telescope, flew aboard the space platform {\it ASTRO-SPAS\/}
during the 1996 November/December mission of the space shuttle {\it Columbia}.
The \orf\ project and the {\it ASTRO-SPAS\/} platform are described in
\citeN{Getal91}, while performance of the Berkeley
spectrometer during this, its second flight,
is discussed in \citeN{Hetal98}.  For reference, the 
effective area peaks at about 9 cm$^2$; the resolution
$\lambda/\Delta \lambda = 3000$ over the 390--1200 \AA\ band.

IC 63 was observed twice, for
a total of 3211 sec.  The target coordinates were
$\alpha$ = 00 59 01.29, $\delta$ = +60 53 17.90 (J2000).
Absolute positioning of the 26 $\arcsec$ diameter
ORFEUS entrance aperture is accurate to $\pm$ 5 $\arcsec$.
This location is coincident with that used by \citeN{WSBB89}
during previous observations with the IUE.
We can find no measurement of the nebular proper motion
in the literature, but if it is comparable
to that of $\gamma$ Cas, the cumulative motion
in the decade since the IUE observations should
be well below one arcsecond.

Detector background is estimated from nonilluminated regions
above and below the dispersed spectrum.  Stray light
creates a high background level between 900 and about 1000 \AA,
limiting the utility of the data for very faint sources
such as IC 63.  Diffuse ``airglow'' (including geocoronal/interplanetary
emission) is scaled from an off-axis aperture displaced by 2.6 $\arcmin$
from the primary aperture. 
Statistical and systematic errors associated with 
background and airglow subtraction are properly tracked.
As a final step, we bin the data on 1 \AA\ centers.

\begin{figure}
\epsscale{.8}
\plotone{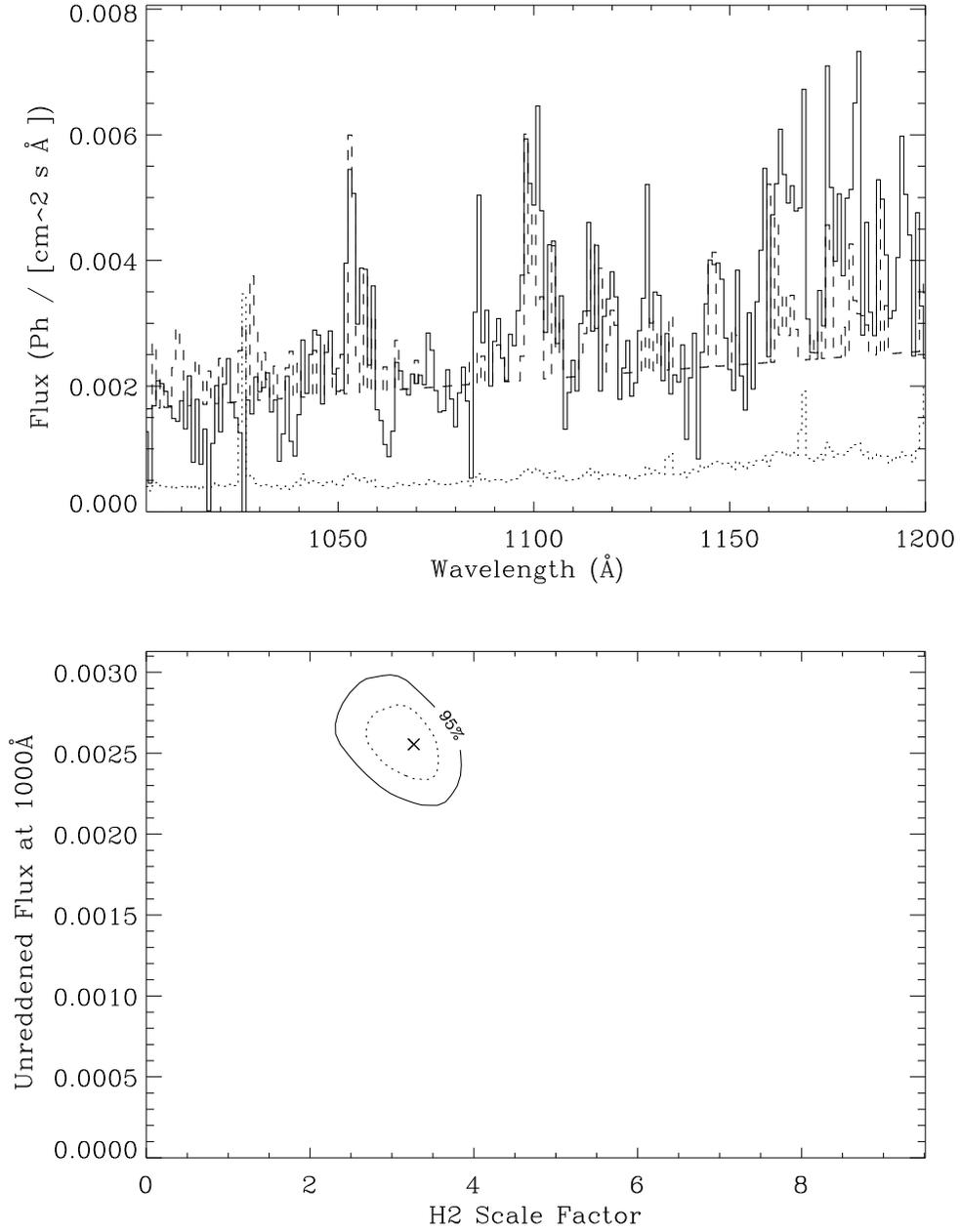}
\caption{Upper panel: Observed spectrum of IC 63 (solid line),
best-fit model (dashed line), and 1 $\sigma$ error (dotted line).
Lower panel: Joint confidence interval on continuum flux
at 1000 \AA\ and \htwo\ scale factor.  See text for definitions.
\label{fluorfig}
}
\end{figure}

The observed spectrum (solid line) and 1 $\sigma$ error (dotted line)
are shown in the upper panel of \fig{fluorfig}.

\section{MODELING OF THE CONTINUUM AND FLUORESCENCE SPECTRUM}
\label{modeling}

To model the observed spectrum, we assume that the
intrinsic emission consists of a linear continuum plus the
reference far ultraviolet \htwo\ fluorescence
spectrum from \citeN{S89}.  
The important physical parameters on which Sternberg's reference 
spectrum is based are: a nebular density of 100 cm$^{-3}$, a temperature
of 100 K, and an incident UV continuum flux of 
$3.4 \times 10^{-6}$ photons cm$^{-2}$ s$^{-1}$ Hz$^{-1}$ at 1000 \AA\
(about 100 $\times$ the typical interstellar value). 
The model predicts that the relative strengths of the UV emission 
features are comparatively unaffected by the detailed parameter values.
The strength of the Werner emission features
relative to those from the Lyman electronic level 
varies somewhat with the shape of the incident radiation field, 
but the dependency is a weak one.
Our software accommodates a fixed foreground reddening,
using the standard interstellar extinction curve (\citeNP{CCM89},
with $R_V$ = 3.1).
The two free parameters of the fit are
the unreddened \htwo\ scale factor,
and the unreddened continuum level at 1000 \AA.
We fix the (reddened) continuum at 1275 \AA\
to the value observed previously
($3.6 \times 10^{-6}$ ergs cm$^{-2}$ s$^{-1}$ sr$^{-1}$ \AA $^{-1}$)
\cite{WSBB89}.
Higher-order polynomial continua did not significantly
improve the quality of the fit.
Our normalization for the unreddened \htwo\ scale factor follows
Sternberg; a scale factor of unity corresponds
to a total UV fluorescent 
flux of $1.1 \times 10^{-4}$ ergs cm$^{-2}$ s$^{-1}$ sr$^{-1}$.

It is important to be cognizant that the sum of the UV 
fluorescent line strengths in the tabulation of \citeN{S89}
does not equal 1000 in the normalized units 
adopted in that work.
The text states that the tabulation represents
the 260 brightest lines, but only after manually coadding the 
line fluxes did we discover that their
summed contribution comprises only 45.6\% of the total radiated power.
In this work we assume that the weaker lines follow
the spectral distribution of the stronger ones,
and simply scale the Sternberg reference spectrum
upward to account for the missing flux, which is
reasonable (Sternberg private communication).
We were not aware of the need for this renormalization
factor during our analysis of the emission from the Taurus 
molecular cloud \cite{H94}.
In that work we noted that the interstellar radiation
field was too faint (by a factor of two)
to excite the observed \htwo\ fluorescence; the
renormalization factor neatly resolves this puzzle.

Returning now to IC 63, we note that the column density of 
interstellar H I toward the nearby illuminating star $\gamma$ Cas
is about $1.5 \times 10^{20}$ cm$^{-2}$ \cite{DS94}.
Foreground molecular gas is negligible (Jenkins, private communication).
With the usual reddening vs. H I relationship \cite{BSD78},
E$_{B-V}$ is about 0.03 magnitudes.
We adopt this foreground reddening for IC 63 as well.
In fitting the model to the data, we exclude a few wavelength
bins where strong gas phase interstellar absorption
is likely to be present, e.g. bins near Lyman $\beta$, 
C II 1036.3 \AA, and N II 1084 \AA.

Our best-fit model is
illustrated in the upper panel of Figure 1 (dashed line).
Detailed parameters and confidence intervals \cite{LMB76}
are illustrated in the lower panel.
The reduced $\chi^2$ is about 1.4 at best fit.

\section{DISCUSSION}
\label{discussion}

The far ultraviolet continuum from IC 63 will
be discussed in a separate work (Gordon \& Hurwitz, in prep).
Here we focus our attention on the \htwo\ fluorescence.

Apart from the region longward of 1170 \AA, where the data 
reveal emission features that we have not been able to identify,
the general appearance of the model fluorescent spectrum
is in good agreement with our observational
data in the 1000 -- 1200 \AA\ region.  
The absolute scaling of the fluorescence, however,
differs significantly from the values found previously.
Witt et al. measured a flux 
of about $7.5 \times 10^{-4}$ ergs cm$^{-2}$ s$^{-1}$ sr$^{-1}$
between 1350 and 1650 \AA.
Adopting the line strengths from Sternberg, we find
that the IUE band encompasses 26\% of the radiated
UV power, leading to a total UV flux of 
about $2.9 \times 10^{-3}$ ergs cm$^{-2}$ s$^{-1}$ sr$^{-1}$,
or equivalently, an \htwo\ scale factor of 26.

This is an observational limit based on what we will
refer to as the IUE-band flux, and includes the effects
of reddening.  For comparison with our dereddened constraints
shown in Figure 1, we must remember to deredden the IUE
band as well, leading to an \htwo\ scale factor
of about 31 (if E$_{B-V}$ = 0.03) in the IUE band.
We find a best fit \htwo\ scale factor of about 3.3 in the
ORFEUS band, and a 95\% confidence range from 2.2 to 3.9 .
Our measurements fall an order of magnitude or more below 
the IUE-band results.

Interestingly, there was some evidence for a shortfall in
the measured flux compared to the fluorescence model
within the IUE spectrum.  
In Figure 2 of \citeN{WSBB89} it can be seen that the data 
are below the model at wavelengths below about 1300 \AA.
The general IUE flux level at $\sim$ 1200 \AA\ 
would produce about 0.0025 ph cm$^{-2}$ s$^{-1}$ \AA $^{-1}$
in the ORFEUS aperture, in good agreement with our own measurements.
The faintness of the short wavelength IUE data has engendered
little previous commentary, perhaps because of the perception that 
correction for Lyman $\alpha$ emission and/or stray starlight 
introduced too much uncertainty at the shortest IUE wavelengths.

Most of the IUE spectrum is dominated by Lyman band transitions 
to high vibrational states.
The ORFEUS spectrum includes both Lyman and Werner features;
the latter generally form the most prominent complexes.
We have fit the ORFEUS spectrum with varying ratios of
Lyman and Werner band line fluxes, but the goodness of fit 
does not improve measurably.  Gross distortions (e.g.
complete exclusion of the Werner band) fit the data very poorly,
and even at this extremum the ORFEUS spectrum requires
an \htwo\ scale factor no greater than 6.

The IUE aperture (10 $\times$ 20 $\arcsec$) 
was somewhat smaller than that used with ORFEUS.  
Since neither instrument provides useful spatial information
within the field of view, both might
underestimate the true specific intensity
if the fluorescence region is very centrally concentrated.
Our larger aperture makes us more susceptible to dilution effects,
which could introduce a factor of up to 2.6 in the
comparison of our results with those of IUE.
We note however that other tracers of central
concentration in IC 63 would not predict a strong fall-off in 
the fluorescent intensity until significantly larger 
angular scales are reached (\citeNP{LLBJF97}; \citeNP{JVDB94}).
If the general IR-to-UV ratio is comparable to that predicted by Sternberg,
the measurements of \citeN{LLBJF97} suggest that the dilution
within their 74$\arcsec$ beam was no worse
than a factor of $\sim$4, e.g. that the fluorescent
region is unlikely to be much smaller than 30 or 40$\arcsec$ 
in diameter.

An increase in the dust extinction to which the
fluorescent spectrum is subjected could
preferentially attenuate the shorter wavelengths, 
helping to reconcile our results with those of \citeANP{WSBB89}.
Complete reconciliation of the ORFEUS and IUE-band 
by this effect would require roughly 1.5 magnitudes
of visual extinction. 
Given the low extinction toward $\gamma$ Cas,
general interstellar dust is unlikely to be responsible.
The nebula itself contains an ample supply of dust; 
significant differential extinction could arise if the
fluorescence region is hidden behind gas
that is presumably in atomic form.  
There are two difficulties with this scenario.
The absorbing slab would increase the
unreddened \htwo\ scale factor to about 130, requiring 
an incident radiation field well in excess of
that from $\gamma$ Cas.  
Furthermore the infrared fluorescent line 
intensities (which are far less affected by extinction)
would be much brighter than those observed \cite{LLBJF97}.

The \htwo\ emission spectrum from 1140 to 1680 \AA\ 
excited by 100 eV electron impact has recently been 
published \cite{LAMJA95}.
J. Ajello kindly provided these data, and the spectrum 
from 1200 to 1720 \AA\ produced by 16 eV electrons, in electronic format.
The higher energy electron impact spectrum shares some
overlap with the ORFEUS spectral band, but the observational
data do not enable us to distinguish this
process from ordinary photon fluorescence.
The ``hardness ratio,'' here defined as the 1220 -- 1300 \AA\
flux divided by the flux longward of 1400 \AA,
is comparable for the photon fluorescence spectrum
and the 16 eV electron impact spectrum.  The 100 eV
spectrum is about 50\% harder.
The electron excitation mechanism thus does not appear 
to offer a great advantage in reconciling the ORFEUS spectrum
with that of IUE.  Extension of the electron impact spectra to shorter
wavelengths (and probably to lower electron energies) would
nonetheless be of utility.

The potential for absorption from
quiescent \htwo\ seems plausible in this environment,
but its signature (a series of bands at wavelengths
below 1108 \AA) is not evident in the spectrum.
All of the fluorescent features in the ORFEUS band,
not merely those below 1108\AA, are weak compared to the IUE lines.

\citeN{WSBB89} predicted that the fluorescent emission
at shorter wavelengths would be fainter than the
model predictions, invoking 
optical depth effects within the fluorescent zone itself.
The model of Sternberg includes the effect of dust
and gas on the incident radiation field,
and properly treats dust extinction as the fluorescent 
photons exit the cloud.  It is ``optically thin,''
however, in that the outgoing radiation is not 
subjected to absorption from the rovibrationally excited \htwo,
a process that becomes increasingly important at shorter wavelengths.
Even if the interaction is primarily a resonant scattering,
photon trapping will burden the shorter wavelength photons
with a longer effective path length and therefore
a higher effective extinction before they can escape.
This mechanism should convert a
fraction of the stellar radiation field
into an ``excess'' of infrared emission
(compared to models where the effect is not included).
The observed infrared emission from IC 63 exceeds the 
total incident stellar power \cite{JVDB94},
so we can say only that the data do not rule
out the possibility of this reprocessing effect.

\section{CONCLUSIONS}
\label{conclusions}

We have detected \htwo\ fluorescence at 
wavelengths between 1000 and 1200 \AA\ from the 
bright reflection nebula IC 63 with the Berkeley
spectrograph aboard the \orf-II mission \cite{Hetal98}.
The wavelengths and relative strengths of the
fluorescent features within the ORFEUS band
agree well with the predictions of the
model of \citeN{S89}.
The absolute fluorescent intensity is fainter 
than an extrapolation from observations 
at longer ultraviolet wavelengths \cite{WSBB89}
by a factor of ten.
The measurements can not be reconciled by
differential extinction from a foreground slab of dust
(presumably associated with neutral gas)
nor by absorption from quiescent \htwo.
Optical depth effects in the fluorescing \htwo\ itself,
predicted by \citeN{WSBB89}, remain the
most plausible mechanism to explain our observations.

\acknowledgments

We acknowledge our colleagues on the
\orf\ team and the many NASA and DARA personnel who helped make the
\orf-II mission successful. This work is supported by NASA grant
NAG5-696.

\clearpage 




\begin{thebibliography}{}

\bibitem[\protect\citeauthoryear{{Bohlin}, {Savage}, \& {Drake}}{{Bohlin}
  et~al.}{1978}]{BSD78}
{Bohlin}, R.~C., {Savage}, B.~D.,  \& {Drake}, J.~F. 1978, ApJ, 224, 132

\bibitem[\protect\citeauthoryear{{Cardelli}, {Clayton}, \& {Mathis}}{{Cardelli}
  et~al.}{1989}]{CCM89}
{Cardelli}, J., {Clayton}, G.,  \& {Mathis}, J. 1989, ApJ, 345, 245

\bibitem[\protect\citeauthoryear{{Diplas} \& {Savage}}{{Diplas} \&
  {Savage}}{1994}]{DS94}
{Diplas}, A.,  \& {Savage}, B.~D. 1994, ApJS, 93, 211

\bibitem[\protect\citeauthoryear{{Grewing} et~al.}{{Grewing}
  et~al.}{1991}]{Getal91}
{Grewing}, M., et~al. 1991, in Extreme Ultraviolet Astronomy, ed. R.~F.
  {Malina} \& S.~{Bowyer} (Elmsford: Pergammon), 437

\bibitem[\protect\citeauthoryear{{Hurwitz}}{{Hurwitz}}{1994}]{H94}
{Hurwitz}, M. 1994, ApJ, 433, 149

\bibitem[\protect\citeauthoryear{{Hurwitz} et~al.}{{Hurwitz}
  et~al.}{1998}]{Hetal98}
{Hurwitz}, M., et~al. 1998, ApJ

\bibitem[\protect\citeauthoryear{{Jansen}, {van Dishoeck}, \& {Black}}{{Jansen}
  et~al.}{1994}]{JVDB94}
{Jansen}, D.~J., {van Dishoeck}, E.~F.,  \& {Black}, J.~H. 1994, A\&A, 282, 605

\bibitem[\protect\citeauthoryear{{Jansen} et~al.}{{Jansen}
  et~al.}{1995}]{JVDBSS95}
{Jansen}, D.~J., {van Dishoeck}, E.~F., {Black}, J.~H., {Spaans}, M.,  \&
  {Sosin}, C. 1995, A\&A, 302, 223

\bibitem[\protect\citeauthoryear{{Lampton}, {Margon}, \& {Bowyer}}{{Lampton}
  et~al.}{1976}]{LMB76}
{Lampton}, M., {Margon}, B.,  \& {Bowyer}, S. 1976, ApJ, 208, 177

\bibitem[\protect\citeauthoryear{{Liu} et~al.}{{Liu} et~al.}{1995}]{LAMJA95}
{Liu}, X., {Ahmed}, S.~M., {Multari}, R.~A., {James}, G.~K.,  \& {Ajello},
  J.~M. 1995, ApJS, 101, 375

\bibitem[\protect\citeauthoryear{{Luhman} et~al.}{{Luhman}
  et~al.}{1997}]{LLBJF97}
{Luhman}, M.~L., {Luhman}, K.~L., {Benedict}, T., {Jaffe}, D.~T.,  \&
  {Fischer}, J. 1997, ApJ, 480, L133

\bibitem[\protect\citeauthoryear{{Shull} \& {Beckwith}}{{Shull} \&
  {Beckwith}}{1982}]{SB82}
{Shull}, M.,  \& {Beckwith}, S. 1982, ARAA, 20, 163

\bibitem[\protect\citeauthoryear{{Sternberg}}{{Sternberg}}{1988}]{S88}
{Sternberg}, A. 1988, ApJ, 332, 400

\bibitem[\protect\citeauthoryear{{Sternberg}}{{Sternberg}}{1989}]{S89}
{Sternberg}, A. 1989, ApJ, 347, 863

\bibitem[\protect\citeauthoryear{{Witt} et~al.}{{Witt} et~al.}{1989}]{WSBB89}
{Witt}, A.~N., {Stecher}, T.~P., {Boroson}, T.~A.,  \& {Bohlin}, R.~C. 1989,
  ApJ, 336, L21

\end{thebibliography}


\newpage 

\end{document}